\documentclass{emulateapj}
\journalinfo{To Appear in Astronomical Journal}

\slugcomment{}
\shorttitle{X-rays in DK Lacertae}
\shortauthors{D. Takei et al.}

\newcommand{\kt}{\ensuremath{k_{\rm{B}}T}}
\newcommand{\lx}{\ensuremath{L_{\rm{X}}}}

\newcommand{\fx}{\ensuremath{F_{\rm{X}}}}
\newcommand{\nh}{\ensuremath{N_{\rm{H}}}}
\newcommand{\ktl}{\ensuremath{k_{\rm{B}}T_{\rm{L}}}}
\newcommand{\kth}{\ensuremath{k_{\rm{B}}T_{\rm{H}}}}

\begin{document}
\title{Discovery of X-ray Emission in the Old Classical Nova DK Lacertae}
\author{
D.~Takei\altaffilmark{1},
T.~Sakamoto\altaffilmark{2}, \&
J.~J.~Drake\altaffilmark{1}
}
\email{dtakei@head.cfa.harvard.edu}
\altaffiltext{1}{Smithsonian Astrophysical Observatory,
                 60 Garden Street, Cambridge, MA 02138, US}
\altaffiltext{2}{Department of Physics and Mathematics, College of Science and Engineering,
                 Aoyama Gakuin University, 5-10-1 Fuchinobe, Chuo-ku, Sagamihara-shi,
                 Kanagawa 252-5258, Japan}

\begin{abstract}
 We report the discovery of X-ray emission at the position of the old classical nova
 DK Lacertae using the \textit{Swift} satellite. Three observations were conducted
 using the X-ray telescope 62 years after the discovery of the nova, yielding 46
 source signals in an exposure time of 4.8~ks. A background-subtracted count rate was
 9$\pm$2$\times$10$^{-3}$~counts~s$^{-1}$, corresponding to a detection significance
 level of 5$\sigma$. The X-ray spectrum was characterized by a continuum extending up
 to about 7~keV, which can be modeled by a power-law component with a photon index of
 1.4--5.6, or by a thermal bremsstrahlung component with a temperature of 0.7--13.3~keV,
 convolved with interstellar absorption with an equivalent hydrogen column density of
 0.3--2.4$\times$10$^{22}$~cm$^{-2}$. Assuming a distance of 3900~pc to the source, the
 luminosity was 10$^{32}$--10$^{34}$~ergs~s$^{-1}$ in the 0.3--10~keV energy band.
 The origin of X-rays is considered to be either mass accretion on the white dwarf or
 adiabatic shocks in nova ejecta, with the former appearing much more likely. In either
 case, DK Lacertae represents a rare addition to the exclusive club of X-ray emitting
 old novae.
\end{abstract}

\keywords{
novae, cataclysmic variables
---
stars: individual (DK Lacertae)
---
X-rays: stars
}

\section{Introduction}
A nova outburst is a cataclysmic nuclear explosion seen in the later stages of close
binary star evolution. Such a close binary typically comprises a white dwarf and a red
dwarf that fills its Roche lobe and transfers hydrogen-rich gas to its companion. When
the amount of accreted material reaches a critical mass, sudden hydrogen fusion is
triggered by a thermonuclear runaway on the white dwarf surface (e.g.,
\citealt{starrfield2008a}). The released energy and material propagate through the
circumstellar environment with typical ejecta velocities of the order of
1000~km~s$^{-1}$. The mass outflow continues until the nuclear fuel is consumed on the
white dwarf, and the system then reverts back to a quiescent state with accompanying
mass accretion. For extensive reviews of novae see e.g., \cite{warner2003,bode2008c}.

Two different processes can be involved in producing X-rays in quiescence. One of them
is the release of gravitational energy through accretion of material onto the white
dwarf (e.g., \citealt{hernanz2002a}), analogous to other cataclysmic variables (i.e.,
polars, intermediate polars, and dwarf novae). The other is through adiabatic shocks in
the nova ejecta interacting with circumstellar environment, in analogy to a miniature
supernova remnant in time and space scales (e.g., \citealt{balman2005c}). In either
case, X-ray detections of quiescent systems known to have undergone a nova explosion
are quite rare. The new discoveries of X-ray emitting old novae provide us additional
pieces to solve the puzzle of evolutionary channels in binaries followed by different
kinds of cataclysmic variable activity.

Systematic X-ray studies of old novae are important for the purpose. However, existing
wide-field X-ray surveys are not sensitive enough to detect emission from most objects
in quiescence. Dedicated pointed observations are necessary, often with a high risk of
detecting nothing. Here, the \textit{Swift} observatory \citep{gehrels2004} provides an
excellent opportunity to overcome the observational difficulties.  Its X-ray telescope,
the ability of the satellite to obtain snapshot observations with short exposures, and
the Fill-in program to fill in gaps in the planned telescope schedule, provide an
effective means for surveying potentially weak X-ray sources.

In this paper, we report the discovery of X-ray emission from the old classical nova DK
Lacertae (hereafter DK Lac) for the first time. Using the \textit{Swift} satellite, we
detected a weak but significant X-ray signal 62 years after the optical discovery of
its outburst. Despite the limited data with poor photon statistics, we further discuss
these X-ray properties and the nature of DK Lac.

\section{Target (DK Lacertae)}\label{target}
The classical nova DK Lac was discovered on 1950 January 23 at a brightness of 6~mag
from photographic photometry \citep{bertaud1950n}. A peak brightness of 5.8~mag was also
confirmed in data taken a day before the discovery \citep{bertaud1952m}. Photometric and
spectroscopic studies of the nova were subsequently conducted, which continued for about
a year after the discovery (e.g., \citealt{gaposchkin1957t,mclaughlin1950t,walker1951o,
wellmann1951n,ribbe1951t,pohl1951h,bochnivek1951l,lareeon1953t,larsson1954t}). The time
evolution of the nova was moderately fast, with decline rates of $t_{2}$ $\sim$ 19~d and
$t_{3}$ $\sim$ 32~d, where $t_{2}$ and $t_{3}$ are the time to fade by 2 and 3~mag from
the optical maximum, respectively. \cite{duerbeck1981} derived a distance to the source
of 1500$\pm$200~pc based on a maximum magnitude versus rate-of-decline method.

DK Lac has been studied further in quiescence using ground-based telescopes.
\cite{henden2001d} found a slow optical decline began in 2000 September (51 years after
the discovery) for the first time, where the \textit{V}-band magnitudes faded from 16.8
to 19.4~mag by 2001 December (52 years after the discovery). \cite{honeycutt2011}
reported extensive photometric observations in 1990--2009 and witnessed a low brightness
state by about 2~mag in 2001--2003 (51--53 years after the discovery). \cite{katysheva2007m}
found 0.1--0.2~mag oscillations with 0.1~d time-scales in \textit{V}-band monitoring
during 2003 (53 years after the discovery). The binary system was expected to have a
mainly face-on inclination, with $i$ $\lesssim$ 25--50~deg based on H$\alpha$ equivalent
widths and \textit{V}-band magnitudes \citep{honeycutt2011}. The counterpart
of the nova is located at (R.\,A., Decl.) $=$ (22$^{\rm{h}}$49$^{\rm{m}}$46$\fs$91,
$+$53$\arcdeg$17$\arcmin$19$\farcm$3) in the equinox J2000.0 with 0.3$\arcsec$ position
accuracy in the two micron all sky survey catalog (2MASS: \citealt{cutri2003t}).

A spatially-resolved nova remnant was first reported by \citet{cohen1985n}, based on a
marginal detection in 6570~\AA\ and H$\alpha$ images obtained in 1983--1984 (35 years
after the discovery). The source radius was estimated to be 2$\arcsec$ by stellar
profile comparison. A remnant expansion velocity was found to be 1075~km~s$^{-1}$ from
grating spectroscopy in 1984 August \citep{cohen1985n}.  \cite{slavin1995a} also
confirmed the spatially extended emission in a H$\alpha+$[\ion{N}{2}] image (6569~\AA)
from 1993 September (44 years after the discovery), in which the remnant radius was
2.0--2.5$\arcsec$. The distance to the source was revised to 3900$\pm$500~pc based on
nebular expansion parallax \citep{slavin1995a}. \cite{gill2000h} visited the system with
the Wide Field Planetary Camera 2 onboard the Hubble Space Telescope in 1999 March 28
(49 years after the discovery), but no extended H$\alpha+$[\ion{N}{2}] emitting remnant
was detected. At that time, the 3$\sigma$ upper limit to the surface brightness was
found to be 6$\times$10$^{-16}$~erg~cm$^{-2}$~s$^{-1}$~arcsec$^{-2}$.

\section{Observations and Reduction}\label{observations}
We conducted a sequence of three pointing observations at the DK Lac position using the
\textit{Swift} observatory on 2012 April 13, 15, and 17 (Observation ID $=$ 00045888001,
00045888002, and 00045888003, respectively). \textit{Swift} has three instruments in
operation \citep{gehrels2004}: the Burst Alert Telescope (BAT: \citealt{barthelmy2005}),
the X-Ray Telescope (XRT: \citealt{burrows2005t}), and the Ultraviolet and Optical
Telescope (UVOT: \citealt{roming2005t}). Here, we concentrate on the XRT data that show
a significant X-ray detection at the nova position.

The XRT was operated in the state that selects the clocking mode automatically depending
on source fluxes. The data presented here were mainly taken with Photon Counting mode,
which provides full imaging and spectroscopic resolution with a frame time of 2.5~s. For
data reduction, we used the High Energy Astrophysics Software package version 6.11.1 and
the calibration database version x20110725 and m20120206. The data were processed to
level 2 event files with the standard screening criteria. The net total exposure time
($t_{\rm{exp}}$) is 4.8~ks.

\section{Analysis}\label{analysis}
\subsection{Image Analysis}\label{analysis_image}

\begin{figure}[tb]
 \epsscale{1.10}
 \plotone{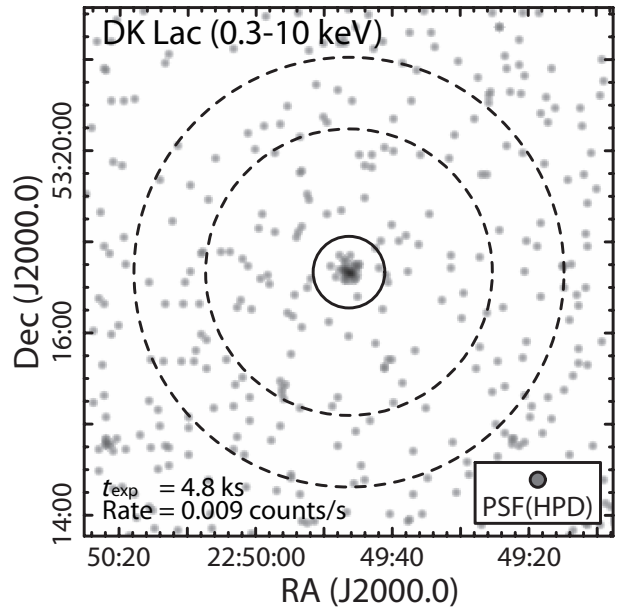}
 \caption{Smoothed XRT image in the 0.3--10~keV energy band. The source and background
 events for DK Lac were accumulated from the solid and dashed areas, respectively. The
 half-power size of the telescope PSF is shown in the bottom-right inset.
 }\label{figure:xrt_image}
\end{figure}

\begin{figure}[tb]
 \epsscale{1.10}
 \plotone{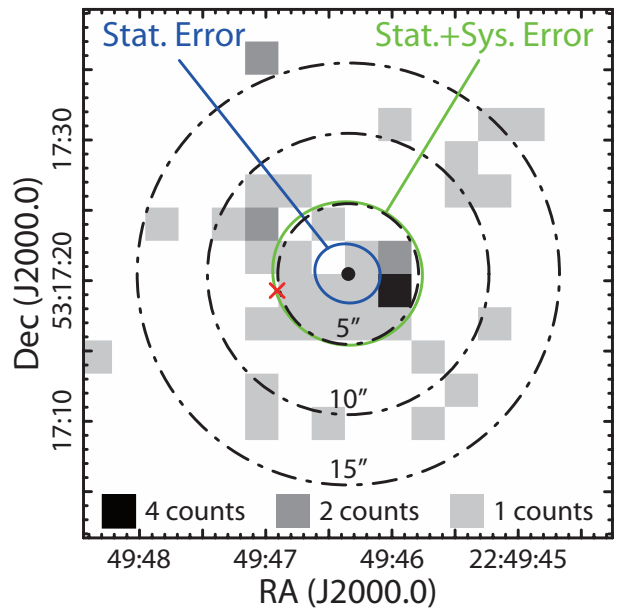}
 \caption{Close-up pixel image around DK Lac in the 0.3--10~keV energy band. The 2MASS
 counterpart position is indicated by the red cross, while the estimated source position
 is shown by a black filled circle. The black dashed-and-dotted circles are distance
 indicators of 5$\arcsec$, 10$\arcsec$, and 15$\arcsec$ from the center. The blue circle
 is a statistical error region of an estimated position. The error region is expand to
 the green circle correcting for the nominal systematic uncertainty of the pointing
 accuracy of \textit{Swift}.
 }\label{figure:position}
\end{figure}

At first data taken with the three \textit{Swift} observations were merged.
Figure~\ref{figure:xrt_image} shows a smoothed image of the resulting XRT data in the
0.3--10~keV energy band. An X-ray source is clearly seen in this image. We estimated the
source position using a two-dimensional fitting procedure with the telescope point
spread function (PSF). The radial profile of the XRT PSF can be numerically approximated
by a King model, $\rm{PSF}(r)$ $=$ $(1+(r/r_{c})^{2})^{-\beta}$, where $r$ is a radial
distance, $r_{c}$ $\sim$ 5$\farcs$8 (2.46~pixels for the CCD detector) is a measure of
the radial scale size, and $\beta$ $\sim$ 1.55 is a slope \citep{moretti2005i}. Events
were taken from a circle with a radius of 20 pixels at positions derived using a
recursive clipping algorithm. The Cash statistic \citep{cash1979p}, a maximum likelihood
function based on the Poisson distribution, was employed to find the best match between
the model PSF and observed source count distribution. The best-fit position is (R.\,A.,
Decl.) $=$ (22$^{\rm{h}}$49$^{\rm{m}}$46$\fs$346, $+$53$\arcdeg$17$\arcmin$20$\farcm$48)
with a 90\% statistical confidence error of $\sim$2$\farcs$3 (see black filled and blue
open circles in Figure~\ref{figure:position}). The 2MASS position of DK Lac is located
at a distance of 5$\farcs$2$\pm$0$\farcs$3, which is consistent with the derived source
position after allowing for the XRT pointing accuracy of 3$\arcsec$ \citep{moretti2006a}.

A total of 46 source events were accumulated from a circle with a radius of 20~pixels,
while 85 background events were taken from an annulus region with inner and outer radii
of 80 and 120 pixels, respectively, at the best-fit position. The background region is
about 20 times larger in area than that of the source, and the background-subtracted
count rate was estimated to be 9$\pm$2$\times$10$^{-3}$~counts~s$^{-1}$ in the 4.8~ks
exposure time, corresponding to a detection significance of 5$\sigma$.

\subsection{Spatial Distribution}\label{analysis_spatial}

\begin{figure}[tb]
 \epsscale{1.10}
 \plotone{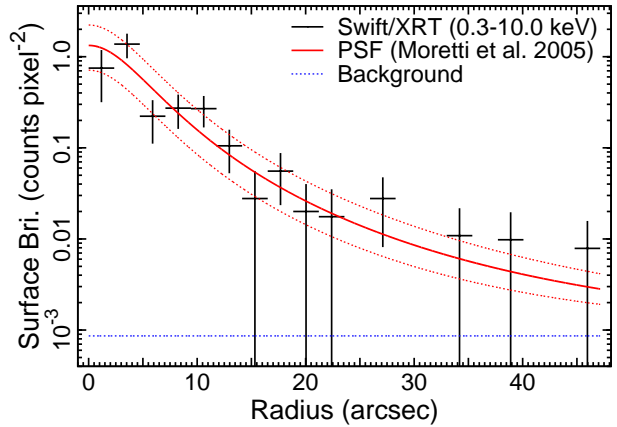}
 \caption{XRT radial profile of DK Lac. Data and 1$\sigma$ errors are shown with black
 crosses. The best-fit PSF model \citep{moretti2005i} and the 1$\sigma$ confidence
 levels are shown with red solid and dashed lines, respectively. A background model is
 by the blue line.
 }\label{figure:dklac_radial}
\end{figure}

\begin{figure}[tb]
 \epsscale{1.10}
 \plotone{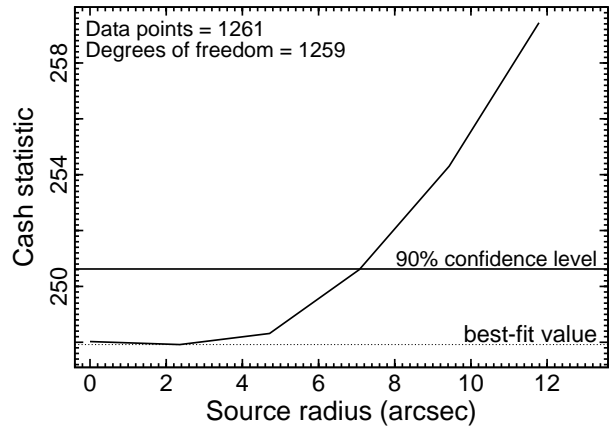}
 \caption{Confidence map of the two-dimensional image fitting to a flat disk model with
 different source radii. The vertical and horizontal axes are the Cash statistic value
 and the source radius, respectively. The two horizontal lines are the best-fit value
 and the upper boundary for the 90\% confidence level, respectively.
 }\label{figure:dklac_conf}
\end{figure}

In order to investigate the possibility that the XRT source has significant spatial
extension, we computed the radial intensity profile at the derived best-fit position
(Figure~\ref{figure:dklac_radial}). The radial profile within a radius of 20 pixels
(47$\farcm$1) was further fitted by a normalized XRT PSF model \citep{moretti2005i} and
an additional background constant, yielding a null hypothesis probability close to
unity and indicating statistical consistency with a point source. An upper limit to the
X-ray source radius was estimated by two-dimensional image fitting using a simple disk
model described by
\begin{equation}
 \rm{Disk}(r) =
  \left\{
   \begin{array}{cl}
    A & (r \le r_{\rm{disk}}) \\
    0 & (r > r_{\rm{disk}})
   \end{array}
   \hspace{1.00mm} [counts],
  \right.
\end{equation}
where an amplitude ($A$) and a radius ($r_{\rm{disk}}$) of an emitting disk are treated
as free parameters. The input source distribution was convolved with the telescope PSF
function \citep{moretti2005i}, and an additional constant parameter was included in the
fitting to represent background.  The data were found to be consistent with a source
radius smaller that $\sim$7$\arcsec$, which corresponds to the 90\% confidence level
based on the Cash statistic (Figure~\ref{figure:dklac_conf}).  The best-fit radius is
about $2\arcsec$, but is also statistically consistent with a radius of zero. The
\textit{Swift} data thus do not allow us to  perform a definitive test of spatial
extent, and are consistent with both a point-like nature and with significant extension
of up to several arcseconds.

\subsection{Temporal and Spectral Analysis}\label{analysis_spectra}

\begin{figure}[tb]
 \epsscale{1.10}
 \plotone{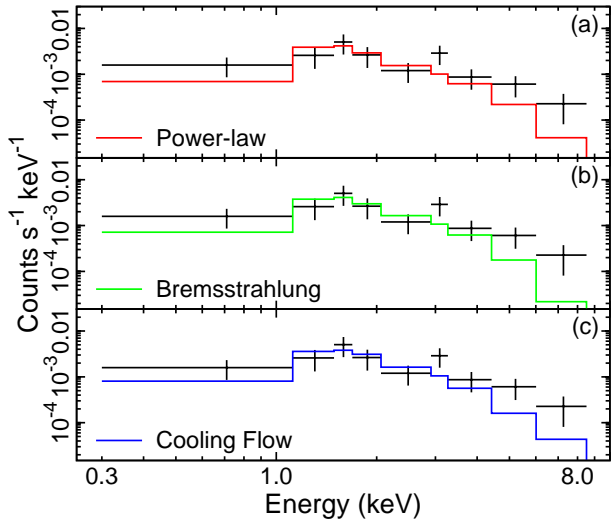}
 \caption{Background-subtracted XRT spectrum of DK Lac with the best-fit (a) power-law,
 (b) bremsstrahlung, and (c) cooling flow models. The data are shown as black crosses,
 while the best-fit models are represented color-coded by solid lines.
 }\label{figure:spectrum}
\end{figure}

We examined the XRT light curve of the source using the same regions as in the above
image analysis, and found the data to be consistent with the assumption of a constant
flux.  We then extracted the background-subtracted XRT spectrum, which is characterized
by a continuum extending up to about 7~keV (Figure~\ref{figure:spectrum}). No
discrete features were found either because of the poor photon statistics or because of
the real spectrum. We fitted an unbinned spectrum in the 0.3--10~keV energy band using
the Cash statistic with a power-law or thermal bremsstrahlung component convolved with
an interstellar absorption model (TBabs; \citealt{wilms2000o}). The best-fit values of
power-law photon index ($\Gamma$), thermal bremsstrahlung temperature ($\kt$), and a
hydrogen equivalent column density ($\nh$) are summarized in Table~\ref{table:spectra}.
The source flux is 10$^{-12}$--10$^{-14}$~ergs~cm$^{-2}$~s$^{-1}$ in the 0.3--10~keV
energy band, which corresponds to an absorption-corrected X-ray luminosity of about
10$^{32}$--10$^{34}$~ergs~s$^{-1}$, assuming a distance of 3900~pc \citep{slavin1995a}.

The X-ray spectrum was also modeled using a cooling flow model originally developed for
clusters of galaxies \citep{mushotzky1988e}, but also known to describe emission from
some cataclysmic variables \citep{mukai2003t,pandel2005x,matranga2012d} . In this model,
the MEKAL code \citep{mewe1985c} is employed to calculate the X-ray spectrum of
optically-thin thermal plasma with a differential emission measure distribution
corresponding to an isobaric radiatively cooling flow. Two dominant parameters, the
higher plasma temperature ($\kth$) and the mass accretion rate ($\dot{M}$), were thawed
for the spectral fitting, while the lower plasma temperature ($\ktl$) and element
abundances were fixed to 0.1~keV and their solar values, respectively.  In this model,
the normalization of the mass accretion rate is also controlled by the redshift value
($z$) representing the source distance. We thus adopted 9.1$\times$10$^{-7}$ for this
parameter, corresponding to the distance to DK~Lac of 3900~pc, assuming the Hubble
constant of 70~km~s$^{-1}$~Mpc$^{-1}$ (e.g., \citealt{jarosik2011s}). The interstellar
absorption (modeled using the TBabs model; \citealt{wilms2000o}) was also treated as a
free parameter. The fitting results obtained with the cooling flow model are also
summarized in Table~\ref{table:spectra}.

\begin{table}[tb]
 \vspace{-3.50mm}
 \begin{center}
  \caption{Best-fit parameters of the X-ray spectrum.}\label{table:spectra}
  \begin{tabular}{llllll}
   \tableline
   Comp.          & Par.                       & Unit                      & Value\tablenotemark{a}               \\
   \tableline
   Absorption     & \nh                        & (cm$^{-2}$)               & 1.1$_{-0.8}^{+1.3}\times$10$^{22}$   \\
   Power-law      & $\Gamma$                   &                           & 2.8$_{-1.4}^{+2.8}$                  \\
                  & \fx\tablenotemark{b}       & (ergs~s$^{-1}$~cm$^{-2}$) & 3.0$_{-2.4}^{+26.7}\times$10$^{-13}$ \\
                  & \lx\tablenotemark{b,c}     & (ergs~s$^{-1}$)           & 3.0$_{-2.3}^{+25.9}\times$10$^{33}$  \\
   \tableline
   \multicolumn{3}{l}{Cash/d.o.f.}                                         & \multicolumn{1}{l}{224.38/966}       \\
   \tableline
   Absorption     & \nh                        & (cm$^{-2}$)               & 0.9$_{-0.6}^{+0.9}\times$10$^{22}$   \\
   Bremsstrahlung & \kt                        & (keV)                     & 2.0$_{-1.3}^{+11.3}$                 \\
                  & \fx\tablenotemark{b}       & (ergs~s$^{-1}$~cm$^{-2}$) & 2.7$_{-1.8}^{+15.0}\times$10$^{-13}$ \\
                  & \lx\tablenotemark{b,c}     & (ergs~s$^{-1}$)           & 1.3$_{-0.8}^{+6.8}\times$10$^{33}$   \\
   \tableline
   \multicolumn{3}{l}{Cash/d.o.f.}                                         & \multicolumn{1}{l}{223.65/966}       \\
   \tableline
   Absorption     & \nh                        & (cm$^{-2}$)               & 1.4$_{-0.7}^{+0.8}\times$10$^{22}$   \\
   Cooling Flow   & \ktl                       & (keV)                     & 0.1 (fixed)                          \\
                  & \kth                       & (keV)                     & 2.6$_{-1.5}^{+8.0}$                  \\
                  & $z$\tablenotemark{c}       &                           & 9.1$\times$10$^{-7}$ (fixed)         \\
                  & $\dot{M}$\tablenotemark{c} & ($M_{\odot}$~yr$^{-1}$)   & 5.0$_{-4.4}^{+39.8}\times$10$^{-9}$  \\
                  & \fx\tablenotemark{b}       & (ergs~s$^{-1}$~cm$^{-2}$) & 2.8$_{-2.5}^{+22.4}\times$10$^{-13}$ \\
                  & \lx\tablenotemark{b,c}     & (ergs~s$^{-1}$)           & 2.2$_{-2.0}^{+17.7}\times$10$^{33}$  \\
   \tableline
   \multicolumn{3}{l}{Cash/d.o.f.}                                         & \multicolumn{1}{l}{220.51/966}       \\
   \tableline
  \end{tabular}
  \begin{flushleft}
   \vspace{-2.00mm}
   $^{\rm{a}}${Statistical uncertainties indicate the 90\% confidence ranges.} \\
   $^{\rm{b}}${Values are in the 0.3--10~keV energy band.} \\
   $^{\rm{c}}${Values are for a distance of 3900~pc.}
   \vspace{-1.50mm}
  \end{flushleft}
 \end{center}
\end{table}

\section{Discussion}\label{discussion}
\subsection{Nature of the X-ray Source}

Based on astrometric consistency, the X-ray emission detected by the \textit{Swift} XRT
is considered to originate from the DK Lac system that underwent a classical nova
explosion in 1950. We derived absorption with an equivalent hydrogen column density of
0.3--2.4$\times$10$^{22}$~cm$^{-2}$ along the line of sight to DK Lac. The values are
consistent with or slightly higher than that of 3$\times$10$^{21}$~cm$^{-2}$ estimated
from HI maps \citep{dickey1990h,kalberla2005t} for a 1$\arcdeg$ cone radius around the
DK~Lac position. The absorption uncertainties depend on the adopted models, and the data
preclude any indication of whether there might be excess absorption related to the
circumstellar medium.

High energy emission in young and old novae can be divided into five known classes: (1)
enigmatic non-thermal components \citep{takei2009s,abdo2010g}, (2) thermal plasma due
to wind shocks in ejecta (e.g., \citealt{mukai2001t,tsujimoto2009a}), (3) photospheric
supersoft sources (e.g., \citealt{takei2008,schwarz2011s}), (4) rekindled accretion
(e.g., \citealt{hernanz2002a,takei2009s}), and (5) extended remnant emission (e.g.,
\citealt{balman2005c}). The former three cases are typically seen during a nova event,
and the remaining two in quiescence. The 1950 outburst of DK Lac was classified as a
moderately fast nova, in which the event had almost terminated one year after its
discovery \citep{bertaud1952m}. Thus the origin of X-rays 62 years after discovery is
considered to be one of the latter cases noted above, i.e., due to rekindled accretion
or nova remnant emission.

\subsection{Rekindled Accretion}
One of the favored scenarios for the X-ray emission in quiescence is accretion on the
white dwarf. A nova event could temporary terminate the mass accretion process through
destruction of an accretion disk during the blast, and by outward hydrodynamic pressure
from a strong radiatively-driven wind \citep[e.g.][]{drake2010a}. The accretion process
would then be expected to rejuvenate over time. Resumption of accretion soon after the
blast has been suggested for e.g., the recurrent nova U~Sco at 10 days after outburst
\citep{mason2012u}, the classical nova V2491~Cyg at 50 days \citep{takei2009s}.
\citet{hernanz2002a} found signs of accretion in V2487~Oph 2.7 years after that outburst.
X-rays from rekindled accretion in classical novae also have been indicated for about
20 objects using the \textit{HEAO-1} satellite \citep{cordova1981t}, the \textit{ROSAT}
\citep{orio2001x}, and more recent observations with \textit{ASCA}, \textit{RXTE},
\textit{Suzaku}, \textit{Chandra} and \textit{XMM-Newton} (e.g.,
\citealt{mukai2005,orio2009n}). Thus, rekindled accretion in DK Lac 62 years after
discovery would not be an unusual scenario. The large flickerings in optical brightness,
presumably from accretion activity (e.g., \citealt{honeycutt2011}), also support the
interpretation that the process was rejuvenated.

For a nova explosion, the mass accretion rate of the binary system should be lower than
$\sim$10$^{-7}$~$M_{\odot}$~yr$^{-1}$ to cause a thermonuclear runaway on a white dwarf
surface without steady nuclear burning during accretion \citep{nomoto1982a}.
The derived accretion rate of $\lesssim$5$\times$10$^{-8}$~$M_{\odot}$~yr$^{-1}$ is
small enough to engender a nova explosion.  The time interval between nova eruptions is
roughly in the range 10--10$^{6}$~yr or so, and mainly depends on the white dwarf mass
and accretion rate (e.g., \citealt{yaron2005a,starrfield2008a}).

The nature of mass accretion in DK Lac depends on the magnetic field strength of the
white dwarf.  For non-magnetic, or weakly magnetized, white dwarfs, an accretion disk
is formed in the system, as in the dwarf nova case. For magnetized white dwarfs, the
trajectory of infalling material is governed by the magnetic field, and an accretion
funnel is formed around a magnetic pole on the white dwarf surface, as in the case of
polars and intermediate polars \citep{cropper1990t,patterson1994t}. The estimated
temperature of the X-ray emitting region of DK Lac ($\lesssim$10~keV) is slightly lower
than the range of those in magnetized systems (e.g., \citealt{yuasa2010w}), but is
similar to those in dwarf novae (e.g., \citealt{pandel2005x,ishida2009s}). The X-ray
luminosities are also consistent with those in other non-magnetic or weakly magnetized
cataclysmic variables (e.g., \citealt{baskill2005t}). Although the large uncertainties
of these values do not allow us to deduce the magnetic type exactly, the X-ray emission
from DK Lac in quiescence represent most closely that expected from rekindled accretion
in dwarf novae.

\medskip
\subsection{Nova Remnant}
The other possible scenario is that X-ray emission originates from the surrounding nova
remnant. The energy and material released in the nova explosion propagate through the
circumstellar environment, and are expected to form a shock structure similar to those
in supernova remnants. From this analogy, if we can observe nova remnants with
sufficient sensitivity before the remnant energy has dissipated, we would naturally
expect X-ray emission produced by adiabatic shocks.

X-rays from nova remnants have only been reported in three objects to date: GK~Per,
RS~Oph, and T~Pyx. GK~Per has the largest known remnant in solid angle with a radius
of about 50$\arcsec$ (4$\times$10$^{12}$~km at a distance of 470~pc), in which knots
and clumps were spatially resolved in X-ray images obtained about 100 years after
the outburst \citep{balman1999t,balman2005c}. An X-ray nebular spectrum was observed,
characterized by a combination of a non-equilibrium ionization thermal plasma model and
a power-law component convolved with photoelectric absorption. The plasma temperature
and photon index were approximately 1~keV and 2, respectively, and the X-ray luminosity
was estimated to be $\sim$10$^{31}$--10$^{33}$~ergs~s$^{-1}$ \citep{balman2005c}. While
GK~Per is a system for which a nova outburst has been observed only once in a human time
scale, T~Pyx and RS~Oph are known to show such outbursts recurrently. The X-ray
properties of T~Pyx and RS~Oph are otherwise fairly similar to those of GK~Per: in all
cases the presence of dense circumstellar material is expected to produce a high
temperature plasma through strong shocks. The reported extended nebulae of T~Pyx and
RS~Oph in X-rays were much more compact in solid angle than that of GK~Per, with the
radii of $\lesssim$10$\arcsec$ \citep{luna2009c,balman2010t}. In the case of T~Pyx,
\citet{Balman2012d} and \citet{Balman2012t} find extension of less than 1$\arcsec$
based on \textit{Chandra} observations, though \citet{montez2012n} find no evidence
for extended emission.

We have noted that the current XRT data do not allow us to make a definitive conclusion
regarding the presence of an extended X-ray remnant for DK~Lac. At face value, the
best-fit X-ray source radius is 2$\arcsec$ (Figure~\ref{figure:dklac_conf}). This value
corresponds to the size of the extended optical shell \citep{cohen1985n,slavin1995a},
and also does not exceed the expected size of $\sim$3--4$\arcsec$ at the epochs of our
X-ray observations. It corresponds to a nebular radius of 1$\times$10$^{12}$~km at a
distance of 3900~pc, which is of the same order as that of GK Per. The estimated
temperature and luminosity of DK~Lac are similar to those of the other nova. If DK~Lac
has an extended X-ray remnant with a radius of a few arcsec, it could be confirmed with
the higher spatial resolution of the \textit{Chandra} X-ray Observatory.

There are, however, three opposing arguments to the extended remnant scenario for
explaining the X-ray emission. (1) The difference in optical and X-ray luminosities:
the upper limit of the optical flux 49 years after discovery was
6$\times$10$^{-16}$~ergs~cm$^{-2}$~s$^{-1}$~arcsec$^{-2}$ \citep{gill2000h}, which
corresponds to an optical luminosity of $\sim$10$^{30}$~ergs~s$^{-1}$~arcsec$^{-2}$ at
a distance of 3900~pc. Even if the DK Lac nebular has a radius of 7$\arcsec$ at maximum,
the X-ray luminosity of 10$^{32}$--10$^{34}$~ergs~s$^{-1}$ is slightly higher than that
in the optical. This implies that the dominant nebular emission is in X-rays, which is
not the case for other novae. (2) Currently, there is no clear evidence of radio
emission --- if DK~Lac has a similar nebular to GK~Per, the radio flux is expected to
be about 0.3~mJy in 1.5~GHz at 3900~pc from \citet{seaquist1989a}. This is enough to be
detectable but DK~Lac was not observed in previous radio surveys (e.g.,
\citealt{bode1987r}).
Future observations at radio wavelengths to search for this would be worthwhile.
(3) There is no evidence of dense circumstellar material. The three known novae with
extended X-ray nebulae are all rather different from typical classical novae. GK~Per
had a planetary nebula before the outburst \citep{seaquist1989a}, while T~Pyx and RS~Oph
are recurrent novae that have substantial pre-outburst circumstellar material with which
the ejecta can interact. Neither of these situations are the case for DK~Lac. While we
cannot conclude that a substantial X-ray remnant is not present, it is much more likely
that rekindled accretion is responsible for the observed X-rays.

\medskip
\section{Conclusion}\label{conclusion}
We have discovered X-ray emission from the old classical nova DK Lac, 62 years after
its only known outburst. The emission originates either in mass accretion onto the
white dwarf, or in adiabatic shocks in the nova remnant. The present data do not allow
us to discriminate between these two scenarios, although rekindled accretion is more
likely. Higher resolution X-ray imaging observations to resolve any extended X-ray
emission and radio observations to search for emission resulting from circumstellar
shocks would be of considerable interest. In either case, DK Lac would represent a
rare addition to the currently very small sample of X-ray emitting old novae.

\acknowledgments

The authors thank the \textit{Swift} principal investigator and operations team for
allocating and scheduling the telescope time for our novae survey program. D.\,T. is
financially supported by the Japan Society for the Promotion of Science. J.\,J.\,D.
was supported by the NASA contract NAS8-39073 to the CXC and thanks the Director,
H.\,Tananbaum, for continuing advice and support. Finally, we thank an anonymous
referee for a very useful report that enabled us to significantly improve the
manuscript.


\end{document}